\documentclass[aps,prl,twocolumn,superscriptaddress,floatfix,noeprint,10pt]{revtex4-2}

\usepackage{graphicx,makecell}
\usepackage{amsmath,amsfonts,amssymb}
\usepackage{mathtools}
\usepackage{mathrsfs}
\usepackage{bm,dsfont}
\usepackage[dvipsnames]{xcolor}
\usepackage[colorlinks=true, citecolor=Green, linkcolor=BrickRed, urlcolor=NavyBlue]{hyperref}
\usepackage[all]{hypcap}

\DeclareMathOperator{\sgn}{sgn}

\begin{document}

\title{Flux attachment theory of fractional excitonic insulators}

\author{Steven Gassner}
\affiliation{Department of Physics and Astronomy, University of Pennsylvania, Philadelphia, Pennsylvania 19104, USA}

\author{Ady Stern}
\affiliation{Department of Condensed Matter Physics, Weizmann Institute of Science, Rehovot 7610001, Israel}

\author{C. L. Kane}
\affiliation{Department of Physics and Astronomy, University of Pennsylvania, Philadelphia, Pennsylvania 19104, USA}

\date{\today}

\begin{abstract}
The search for fractional quantized Hall phases in the absence of a magnetic field has primarily targeted flat-band systems that mimic the features of a Landau level. In an alternative approach, the fractional excitonic insulator (FEI) has been proposed as a correlated electron-hole fluid that arises near a band inversion between bands of different angular momentum with strong interactions. It remains an interesting challenge to find Hamiltonians with realistic interactions that stabilize this state. Here, we describe composite boson and composite fermion theories that highlight the importance of $(p_x+ip_y)^m$ excitonic pairing in stabilizing FEIs in a class of band inversion models. We predict a sequence of Jain-like and Laughlin-like FEI states, the simplest of which has the topological order of the bosonic $\nu=1/2$ fractional quantized Hall state. We discuss implications for recent numerical studies on a chiral spin liquid phase in interacting Chern insulator models.
\end{abstract}

\maketitle

\noindent {\color{blue}\emph{Introduction.}}
A recent focus in quantum condensed matter physics has been the search for correlated electronic states that resemble fractional quantized Hall (FQH) states but arise at zero magnetic field.   Fractional Chern insulators (FCIs) were originally proposed theoretically for fractionally filled, nearly flat Bloch energy bands with non-zero Chern number in the presence of strong interactions \cite{tang2011,neupert2011,sun2011,regnault2011,parameswaran2013,liu2013review,neupert2015, PhysRevLett.133.246602, abouelkomsan2024compressiblequantummattervanishing}.   Interest in this paradigm has been heightened by recent experiments reporting FCIs in two dimensional (2D) transition metal dichalcogenide materials \cite{Cai2023,Zheng2023,Park2023}, and multilayer graphene devices \cite{Lu2024}.

The fractionally filled flat Chern band provides a route to a FQH state that is closest in spirit to the fractionally filled Landau level.  Rules of thumb, like optimizing the uniformity of the Berry curvature and saturating the trace condition on the quantum metric, are mechanisms for engineering Bloch bands that mimic a Landau level, which realizes a FCI when fractionally filled \cite{roy2014, claassen2015, ledwith2023}.  
For a pure 2D electron gas, Galilean invariance dictates that the Hall conductivity is tied to the Landau level filling, $\sigma_{xy}=\nu e^2/h$.   Thus, FQH plateaus occur for $\nu \approx \sigma_{xy}h/e^2$ for a particular set of $\nu$'s.   A periodic lattice, however, eliminates the correspondence between $\sigma_{xy}$ and $\nu$ \cite{hofstadter1976,tknn1982,kol1993}.   This poses the question of whether a fractional quantized band filling is essential for realizing a FQH state.   If not, then what determines the fraction?

In recent work, a different route has been considered, in which a FQH state arises in a correlated fluid of electrons and holes, dubbed a fractional excitonic insulator (FEI) \cite{Hu2018}.  A model wavefunction was considered of the form \cite{dubail2015}
$|\Psi_m\rangle = \sum_N (f^N/N!)|\psi_m^N\rangle$,
    \begin{equation}
        \psi^N_m(\left\{z_i,w_j\right\}) = \frac{\prod_{i<i'}(z_i-z_{i'})^m\prod_{j<j'}(w_j-w_{j'})^m}{\prod_{i,j}(z_i-w_j)^m},
        \label{wavefunction}
    \end{equation}
where $z_i$ ($w_j$) represent the complex coordinates of electrons (holes) with fugacity $f$.   This wavefunction resembles a two-component Halperin wavefunction \cite{halperin1983}, but it does not include the Gaussian factor that arises in a Landau level.  It does not require a fractionally quantized band filling, but it does require the densities of electrons and holes to be equal.   It was shown that for $m=1$ Eq. (\ref{wavefunction}) is the exact groundstate of a simple non-interacting two band model for a Chern insulator.   Using a variant of Laughlin's plasma analogy \cite{laughlin1983}, it was argued that for odd integer $m>1$ (\ref{wavefunction}) describes a Laughlin state composed of charge $\pm e$ fermions (bosons) for $m$ odd (even) with $\sigma_{xy}=(1/m)e^2/h$.   

Evidence that such a FEI state is possible included (i) a composite fermion mean field theory, and (ii) a coupled wire construction.   However, neither of these approaches provided guidance for how to realize this state as the ground state of a physically viable Hamiltonian.   In both cases the value of $\sigma_{xy}$ that results depends on the input to the model.   For (i) it depends on the  flux attachment scheme, as well as the composite fermion Hamiltonian, while for (ii) it depends on the form of the wire couplings that specify a solvable limit.   Thus, unlike the FQH states where $\sigma_{xy}$ is specified by the magnetic field (or the Landau level filling), it is not clear from these theories what $\sigma_{xy}$ should be for a given microscopic model.  Ref. \cite{Hu2018} also introduced a third approach:
(iii) an exact Hamiltonian for (\ref{wavefunction}) was constructed using a method introduced in Ref. \cite{kane91}, which involved specific long ranged multi-body interactions.  If those interactions are ad hoc set to zero, the resulting non-interacting fermion problem describes two coupled bands with angular momentum $m$ excitonic pairing.   This led to the suggestion that the FEI state might be realized by the inversion of two bands that differ in angular momentum by $m$, in the presence of suitable interactions.

It remains to be established whether a FEI ground state can be stabilized for inverted bands with a physically viable interaction.   In this paper, we make partial progress in this direction by showing that within a flux attachment theory based on composite bosons \cite{girvin1987,zhang1989,read1989} or composite fermions \cite{Jain1989,halperin1993}, the angular momentum $m$ coupling between the conduction band and the valence band provides an interaction that stabilizes a FQH state.
This goes beyond the previous theory by clarifying the physical mechanism by which the angular momentum $m$ excitonic pairing selects the flux attachment scheme, which determines the FQH conductivity.   

We will begin by developing the composite boson theory.   We will show that for composite bosons with $p$ bound fluxes ($p$ odd) the excitonic pairing term is only effective at lowering the energy and opening a gap when $p=m$.   Thus, the composite boson theory selects $p=m$.  For $m=1$ this theory correctly predicts an integer quantized Hall state with $\sigma_{xy}=e^2/h$.    For odd integer $m>1$ this theory describes a correlated fluid of electrons and holes that resembles a Laughlin state with $\sigma_{xy} = (1/m) e^2/h$. We will next introduce a composite fermion theory, which describes a larger class of states, analogous to the Jain states \cite{Jain1989}.  We will argue that for angular momentum $m$ excitonic pairing, the flux attachment scheme in which $p$ fluxes are attached to the electron ($p$ even) can result in a gapped composite fermion state with Chern number $n=m-p$. We will conclude by discussing the simplest of these new states, a bosonic state with the topological order of the $\nu=1/2$ FQH state, and its implications for recent numerical work on realizing the chiral spin liquid in a Chern insulator in the presence of strong interactions.

\noindent {\color{blue}\emph{Band Inversion Model.}}
We begin with a Hamiltonian   ${\cal H} = {\cal H}_0 + {\cal H}_{\rm pair}$ describing an angular momentum $m$ band inversion, with
\begin{equation}
\begin{split}
      & {\cal H}_0 = \sum_{\bf k} \left(\frac{|{\bf k}|^2}{2m^*}-E_0\right) \left(\Psi_{e,{\bf k}}^\dagger \Psi_{e,{\bf k}} + \Psi_{h,{\bf k}}^\dagger \Psi_{h,{\bf k}}\right) \\
     & {\cal H}_{\rm pair} =\sum_{\bf k} v i^m (k_x + i k_y)^me^{-|{\bf k}|^2d^2/2} \Psi^\dagger_{e,-{\bf k}}\Psi^\dagger_{h,{\bf k}} + \text{h.c.}
\end{split}
\label{bhz}
  \end{equation}
$\Psi_{e,{\bf k}}^\dagger = c^\dagger_{c,{\bf k}}$ creates an electron ($e$) in the conduction band ($c$) and
$\Psi_{h,{\bf k}}^\dagger = c_{v,-{\bf k}}$ creates a hole ($h$) in the valence band ($v$).  The term ${\cal H}_0$ describes $c$ and $v$ bands (both with effective mass $m^*$) that are inverted for $E_0>0$, which is the case we consider from here on.  The term ${\cal H}_{\rm pair}$ describes the lowest-order symmetry-allowed coupling when the $c$ and $v$ bands differ in angular momentum by $\Delta J_z = \hbar m$. It is dubbed the ``pairing" term because it resembles a mean-field order parameter for excitonic pairing, though we emphasize that it represents an intrinsic hybridization between the $c$ and $v$ bands and not an interaction-induced instability. For $m=1$ it describes ``$p+ip$ pairing," which is possible in the presence of broken time reversal symmetry in a material with inversion and/or $C_2$ rotational symmetries. The condition   $m>1$ requires additional symmetry to forbid smaller $m$.   For example, $m=3$ could arise due the inversion of $s$ and $f$ states in a crystal with $C_6$ rotational symmetry \cite{Venderbos2018}. Finally, the parameter $d$ ensures the Hamiltonian is regularized for $|\mathbf{k}|\to \infty$ and $m>1$, with $1/d$ setting the region in momentum space over which the Berry curvature of the hybridized bands is nonzero.  We note that inverted bands with angular momentum $m$ coupling described by ${\cal H}_{\rm pair}$ also arise in low energy theories of rhombohedral $m$-layer graphene \cite{min2008,koshino2010}.

For $E_0>0$, in the absence of interactions, (\ref{bhz}) describes a Chern insulator with $\sigma_{xy} = m e^2/h$.   For
$m=1$ and $d=0$ it describes a regularized Dirac fermion with negative mass \cite{read2000,qi2006,bernevig2006}.   In that case, for $E_0 = m^* v^2/2$, the ground state wavefunction has precisely the form (\ref{wavefunction}) with $f = m^* v/(2\pi)$ when expressed in terms of the real space operators $\Psi_c^\dagger(z)$ and $\Psi_v^\dagger(w)$ acting on the vacuum defined as the filled valence band \cite{Hu2018}. We seek to identify possible phases for $m>1$ in the presence of strong interactions.

\noindent {\color{blue}\emph{Composite Boson Theory.}}
We define composite boson operators by performing a singular gauge transformation that attaches $p$ flux quanta to electrons in the conduction band, and $-p$ flux quanta to holes in the valence band.  Since opposite fluxes are attached to the electrons and holes, which have equal density, the resulting statistical flux will be zero on the average.   While any odd integer $p$ defines a legitimate statistical gauge transformation with this property, we will see below that the choice $p=m$ is dictated by the form of ${\cal H}_{\rm pair}$.

In terms of the electron (hole) composite boson operators $\tilde\Phi_{e(h)}(z=x+iy)$, the fermion operators are
\begin{equation}
    \begin{split}
       \Psi_e^\dagger(z) &= \tilde\Phi_e^\dagger(z) e^{i p \int d^2z' \Theta(z-z') \rho(z')} \\
       \Psi_h^\dagger(w) &= \tilde\Phi_h^\dagger(w) e^{-i p \int d^2w' \Theta(w-w') \rho(w')} \\
    \end{split}
    \label{boson operators}
\end{equation}
where $\rho(z) = \tilde\Phi_e^\dagger(z) \tilde\Phi_e(z) - \tilde\Phi_h^\dagger(z) \tilde\Phi_h(z)$, and $\Theta(z) = \arg z$.  Substituting (\ref{boson operators}) into (\ref{bhz}), we find
\begin{equation}
    {\cal H}_0 =  \frac{|(\nabla - i {\bf a})\tilde\Phi_e|^2 + |(\nabla + i {\bf a})\tilde\Phi_h|^2}{2m^*}  + V(\tilde\Phi_e,\tilde\Phi_h)\\
\label{boson H}
\end{equation}
with 
${\bf a} = p \int d^2z' \nabla\Theta(z-z')\rho(z')$
and 
$V=  -E_0(|\tilde\Phi_e|^2 + |\tilde\Phi_h|^2)+\frac{u}{2} (|\tilde\Phi_e|^4 + |\tilde\Phi_h|^4) - w |\tilde\Phi_e|^2|\tilde\Phi_h|^2 $.
(\ref{boson H}) includes quartic interactions (with $u>w$), which depend on the fermion interactions.  However, this type of theory does not predict $u$, $w$ precisely. The term  $u$ is present even for non-interacting fermions due to the Pauli principle. 

The pairing term requires a more careful treatment, which will be the crux of our analysis.  In real space, 
\begin{equation}
    {\cal H}_{\rm pair} =  \int d^2\delta z F(\delta z) \Psi_e^\dagger(z)\Psi_h^\dagger(z+\delta z) + \text{h.c.}
\end{equation}
with
\begin{equation}
    F(\delta z) = \frac{v }{2\pi d^{2+2m}} \delta z^m e^{-|\delta z|^2/2d^2}.
    \label{regularized F}
\end{equation}
In terms of the composite boson operators,
\begin{equation}
    {\cal H}_{\rm pair} = \int d^2\delta z \tilde F_p(\delta z) \tilde\Phi_e^\dagger(z) \tilde\Phi_h^\dagger(z+\delta z) + \text{h.c.},
\end{equation}
with
\begin{equation}
    \tilde F_p = F(\delta z) e^{- i p \Theta(\delta z)} e^{i p \int d^2z' \rho(z')(\Theta(z-z')-\Theta(z+\delta z-z'))}.
    \label{F transformation}
\end{equation}
The first exponential accounts for the statistical interaction between the electron and the hole.   The second exponential accounts for the interaction with all of the other electrons and holes and can be expanded for small $\delta z$ to be $1 + i \bar a \delta z + ...$, where $\bar a = a_x - i a_y$.   Using (\ref{regularized F}), the leading term of that expansion (zeroth order in $\delta z$) is
\begin{equation}
    \tilde F_p(\delta z) = \frac{v}{2\pi d^{2+2m}}r^m e^{-r^2/2d^2} e^{i(m-p)\theta},
\label{Fp theta}\end{equation}
with $\delta z = r e^{i\theta}$.  This shows that (\ref{boson operators}) shifts the angular momentum of the electron-hole coupling  by $p$.   Importantly, for $p=m$ the composite electrons and holes are paired with angular momentum zero, represented at long wavelengths by
$\tilde F_{m}(\delta z) = \tilde v \delta^2(\delta z)$ with $\tilde v = 2^{m/2}\Gamma(1+m/2) v/d^m$
\begin{equation}
    {\cal H}_{\rm pair} =  \tilde v (\tilde\Phi_e^\dagger \tilde\Phi_h^\dagger + \tilde\Phi_h \tilde\Phi_e).
    \label{Hpair 0}
\end{equation}
If we had chosen $p\ne m$, then the $\theta$ dependence in (\ref{Fp theta}) forbids the $\delta$-function coupling.  To leading order, $\tilde F_p$ will involve derivatives of a $\delta$-function, leading to
\begin{equation}
    {\cal H}_{\rm pair} \propto \left\{\begin{array}{ll} 
    \tilde\Phi_e^\dagger (\partial_z - i\bar a)^{p-m} \tilde\Phi_h^\dagger + \text{h.c.}  & {\rm for} \ \ p> m,\\
    \tilde\Phi_e^\dagger (\partial_{\bar z} - ia)^{m-p} \tilde\Phi_h^\dagger + \text{h.c.} & {\rm for} \ \ p < m.
    \end{array}\right.
    \label{H pair p><1}
\end{equation}
The $a$ dependence (guaranteed by gauge invariance) follows from the higher order terms in the expansion of the second exponential in (\ref{F transformation}).   Comparing (\ref{Hpair 0}) and (\ref{H pair p><1}), the coupling between the composite electrons and holes provided by ${\cal H}_{\rm pair}$ is lowest order in gradients for $p=m$.   This provides a mechanism for selecting the appropriate flux attachment scheme for the composite boson theory.

We next show for $p=m$ that ${\cal H}_{\rm pair}$ locks the composite electron and the composite hole Bose fluids together, allowing the system to take advantage of ${\cal H}_{\rm pair}$ to lower its energy.
It is simplest to adopt a Euclidean Lagrangian formulation, where $\bf a$ is implemented with a Chern Simons term.   The Lagrangian density is 
\begin{equation}
\begin{split}
{\cal L} = 
    &\tilde\Phi_e^\dagger\left(\partial_\tau  - i a_0-iA_0\right)\tilde\Phi_e  + \frac{1}{2m^*}|(\nabla - i{\bf a}-i{\bf A})\tilde\Phi_e|^2\\
   + &\tilde\Phi_h^\dagger\left(\partial_\tau +i a_0+iA_0\right)\tilde\Phi_h  + \frac{1}{2m^*}|(\nabla + i{\bf a}+i{\bf A})\tilde\Phi_h|^2\\
  + &V(\tilde\Phi_e,\tilde\Phi_h) 
  + \tilde v ( \tilde\Phi_e^\dagger \tilde\Phi_h^\dagger + \tilde\Phi_h \tilde\Phi_e )
   + \frac{1}{m}{\cal L}_{CS}[a_\mu], \label{L}
\end{split}
\end{equation}
where we included an external vector potential $A_\mu$, and
\begin{equation}
    {\cal L}_{CS}[a_\mu] = \frac{i}{4\pi} \epsilon_{\mu\nu\lambda} a_\mu \partial_\nu a_\lambda.
\end{equation}
One can check that the previous definition of the dynamical gauge field is recovered from this Lagrangian via the equation of motion for $a_0$. In the absence of $\tilde v$ and coupling to $a_\mu$, ${\cal L}$ describes two Bose fluids for the composite electrons and holes.   Their densities are determined by $E_0$, and the important low energy degrees of freedom are the phases of $\tilde\Phi_{e,h}$.   To proceed, we write $\tilde\Phi_{e/h} = \sqrt{n_{e/h}} e^{i(\varphi_\sigma \pm \varphi_\rho)}$, where $e^{2i\varphi_{\rho(\sigma)}}$ creates a charge $2e$ (neutral composite electron-hole pair).    The action is minimized for $n_e = n_h = \bar n \equiv E_0/(u-w) $.   We next expand in $\delta n_{e,h} = n_{e,h}-\bar n$ to quadratic order and integrate out the massive fluctuations in $\delta n_{e,h}$ (see Appendix A).
We find
${\cal L} = {\cal L}_\sigma + {\cal L}_\rho$ with
\begin{equation}
\begin{split}
    &{\cal L}_\sigma = \frac{(\partial_\tau\varphi_\sigma)^2}{u-w}  + \frac{\bar n}{m^*}(\nabla\varphi_\sigma)^2  + 2 \bar n\tilde v \cos 2\varphi_\sigma  \\
    &{\cal L}_\rho = \frac{(a_0+A_0)^2}{u+w} + \frac{\bar n}{m^*}({\bf a}+{\bf A})^2   + \frac{1}{m}{\cal L}_{CS}[a_\mu].\label{L sigma rho}
\end{split}
\end{equation}
For ${\cal L}_\sigma$, the $\cos 2\varphi_\sigma$ term will lock $\varphi_\sigma$ and gap the neutral sector.
For this it was crucial that $p=m$.   For other $p$, the coupling is given by (\ref{H pair p><1}), which does not gap the neutral sector.  For $\tilde v=0$, the electron and hole charges are independently conserved, and ${\cal L}_\sigma$ describes a gapless Goldstone mode.  This suggests that the choice $p=m$ is natural because it leads to the most stable gapped composite boson mean field theory. Remarkably, the last term in $\mathcal{L}_\sigma$ shows that the energy gain from $v$ scales with the entire density of inverted electron states (see Appendix B).

With the neutral sector being gapped, the remaining low energy degrees of freedom are described by ${\cal L}_\rho$, which exhibits a gap for charge fluctuations.  Integrating out $a_\mu$ then leaves ${\cal L}_\rho = m^{-1}{\cal L}_{CS}[A_\mu]$, which describes a FQH response with $\sigma_{xy} = m^{-1} e^2/h$. This can also be formulated with a 2$\times$2 $K$-matrix in analogy to the quantum Hall ferromagnet (see Appendix C). For $m=1$, the theory correctly describes the exactly solvable free fermion model, which provides motivation for trusting it in the more difficult fractional case.

The composite boson theory provides a simple description of Laughlin-like states. States with other fractional Hall conductivities can be derived in direct analogy to the hierarchy states of the ordinary FQHE \cite{haldane1983,halperin1984,halperin2020,wen2004} by performing further flux attachment on a $p\neq m$ composite boson theory. Some of these states---analogous to the Jain states---are  simpler to describe in terms of a composite fermion theory, which we describe next.

\noindent {\color{blue}\emph{Composite Fermion Theory.}}
We now consider flux attachment with even integer $p$ to define composite fermions. In the context of the FQH and the FCI the mean-field theory of composite fermions maps a problem that is gapless in the absence of interaction---the partially filled Landau level or Chern band of electrons---to a problem that is gapped, in which an integer number of composite fermion Landau levels or mini-bands are filled. The Hall conductivity $\sigma^{CF}_{xy}$ of the composite fermions is then an integer (in units of $e^2/h$), and that of the electrons, $\sigma^e_{xy}$ is fractional, since $(\sigma_{xy}^e)^{-1}=(\sigma_{xy}^{CF})^{-1}+p$. Here, the non-interacting ground state is gapped, and exhibits an integer Hall conductivity. A necessary condition for a composite fermion starting point to be valid is then that the interaction is strong enough to overcome this gap. When that happens, a gap-closing transition would take the system from an integer to a fractional Hall conductivity state. We now analyze the latter.

Within our flux attachment scheme, the composite fermions see no average flux and exhibit an energy gap.   We assume that this gap is sufficiently large that interactions can be ignored.   Different values of $m$ and $p$ then lead to a sequence of FQH states similar to the Jain sequence.   However, unlike the Jain states, the values of $p$ and the Chern number do not identify a specific fractionally quantized filling factor.  The value of $p$ chosen in a given system will depend on the specific interactions, and is outside the scope of the present theory to determine.   

After  flux attachment (\ref{boson operators}), using (\ref{H pair p><1}) the composite fermion ($\tilde\Psi_{e,h}$) Hamiltonian (ignoring interactions) is
\begin{equation}
\begin{split}
      & {\cal H}_{CF} = \sum_{\bf k}\left(\frac{|{\bf k}|^2}{2m^*}-E_0\right) \left(\tilde\Psi_{e,{\bf k}}^\dagger \tilde\Psi_{e,{\bf k}} + \tilde\Psi_{h,{\bf k}}^\dagger \tilde\Psi_{h,{\bf k}}\right)+ \\
     &  \tilde v i^{m-p} (k_x + i \sgn(m-p) k_y)^{|m-p|} \tilde\Psi^\dagger_{e,-{\bf k}}\tilde\Psi^\dagger_{h,{\bf k}} + \text{h.c.}
\end{split}
\label{cf hamiltonian}
\end{equation}  
Eq. (\ref{cf hamiltonian}) has a gap with Chern number $m-p$, which determines $\sigma_{xy}^{CF}$.  After integrating out $\tilde\Psi_{e,h}$, coupled to $a_\mu$ and $A_\mu$, the Lagrangian is
\begin{equation}
    {\cal L} = (m-p){\cal L}_{CS}[A_\mu + a_\mu] + p^{-1} {\cal L}_{CS}[a_\mu],
\end{equation}
and describes a gapped FQH fluid.  Integrating $a_\mu$ then gives a ${\cal L}_{CS}[A_\mu]$ with a coefficient that determines
\begin{equation}
    \sigma_{xy}=\left(\frac{1}{m-p}+p\right)^{-1} \frac{e^2}{h}.
\end{equation}
For $m$ odd, this sequence of possible states includes the state with $\sigma_{xy}=m^{-1} e^2/h $ predicted by the composite boson theory ($p=m-1$) and the non-interacting Chern insulator with $\sigma_{xy} = m e^2/h$ ($p=0$), as well as others. Interestingly, this scenario raises the possibility of a transition from an integer $\sigma_{xy}=me^2/h$ for weak interaction to a fractional $\sigma_{xy}= m^{-1} e^2/h$ when the interaction becomes strong enough to overcome the band gap.

\noindent {\color{blue}\emph{Bosonic Quantum Hall States and Chiral Spin Liquid.}}
The analysis presented above can be applied to a bosonic system of charge $\pm q$ bosons.    
The analysis is the same, except that the composite boson (fermion) theory requires $p$ even (odd).   In particular, the bosonic theory with $m=2$ is expected to support a ground state similar to (\ref{wavefunction}) with $m=2$.   According to the analysis of Ref. \cite{Hu2018}, this state is more robust than higher values of $m$ because the wavefunction describes a quantum Hall fluid even when the pair density (or fugacity, $f$) is small.  The bosonic $m=2$ state is thus an appealing target.  

Here we outline a scenario in which the bosons consist of bound singlet pairs of electrons.   Consider spinful electrons described by the Hamiltonian (\ref{bhz}) with $m=1$.   Without interactions, the inverted state is a Chern insulator with $\sigma_{xy}=2 e^2/h$.   Beginning in the uninverted state $E_0<0$, suppose that there is a strong attractive interaction between $\uparrow$ and $\downarrow$ electrons, so that the low lying charged excitations are bound pairs created by $\Phi_{2e}^\dagger = \Psi_{e\uparrow}^\dagger \Psi_{e\downarrow}^\dagger$ and $\Phi_{2h}^\dagger=\Psi_{h\uparrow}^\dagger \Psi_{h\downarrow}^\dagger$ with binding energy $\Delta E$.  
For $-\Delta E<E_0 < 0$ it will be energetically favorable to create charge $\pm 2e$ bosonic pairs, while there remains a gap for individual electrons and holes.   The pairing term of Eq. (7) then requires a second order process creating a singlet pair of electrons and a singlet pair of holes.   Provided the interactions can be correctly adjusted, this ``bosonic band inversion" will be described by 
\begin{equation}
    \begin{split}
&{\cal H} =    -(E_0+\Delta E)(|\Phi_{2e}|^2 +|\Phi_{2h}|^2) + \frac{|\nabla \Phi_{2e}|^2+|\nabla \Phi_{2h}|^2}{2m^*}    
\\
 + & \frac{u}{2}(|\Phi_{2e}|^4 +|\Phi_{2h}|^4)-  w|\Phi_{2e}\Phi_{2h}|^2 + v( \Phi_{2e}^\dagger \partial_z^2 \Phi_{2h}^\dagger + \text{h.c.})
    \end{split}
    \label{bosonic nu=2}
\end{equation}
The transformation (\ref{boson operators}) to composite bosons $\tilde\Phi_{2e,2h}$ then proceeds with $p=2$, and predicts a strong paired bosonic FQH state with $\sigma_{xy} = (1/2) (2e)^2/h = 2 e^2/h$.

Recent works \cite{kuhlenkamp2024, divic2024a, divic2024b} have studied a triangular lattice Hofstadter-Hubbard (tHH) model with flux $\pi/2$ per triangle.  As a function of the Hubbard $U$, there is a transition from a Chern insulator with $\sigma_{xy}= 2 e^2/h$ to an insulating state identified as the Kalmeyer-Laughlin chiral spin liquid (CSL) \cite{Kalmeyer1987}.  Ref. \cite{divic2024b} argued that this transition occurs in conjunction with a vanishing of the gap for charge $2e$ singlet pairs of electrons, suggesting the possibility of unconventional superconductivity upon doping.   The tHH model is closely related to the model studied here.   It is identical to a Hubbard model on a bipartite square lattice with $s$ ($p_x+i p_y$) orbitals on the A (B) sublattice with nearest neighbor hopping, along with 2nd neighbor hopping on one diagonal. The hopping amplitudes have equal magnitude $t$ and phases determined by the orbital symmetry.   If we in addition include a staggered sublattice potential $E_s$ (with a role similar to $E_0$ in (\ref{bhz})), then for $|E_s| < 6t$ and $U=0$ the system is a Chern insulator, with two spin-degenerate bands each with $\mathcal{C}=1$.  For $|E_s| = 6t$ the system exhibits a band inversion transition identical to (\ref{bhz}) with $m=1$.  

Divic et al. \cite{divic2024b} considered this model for $E_s = 0$, deep in the Chern insulator phase, and found numerical evidence for charge $2e$ singlet pairing upon increasing $U$, consistent with the scenario of a charge gap closing at the transition to the CSL. This appears similar to the bosonic band inversion scenario (\ref{bosonic nu=2}) described above, where a  fluid of bosonic charge $\pm 2e$ electron pairs and hole pairs (that are excitations of a Chern number two insulator) form the strong paired bosonic state with $\sigma_{xy} = -2 e^2/h$.   The resulting CSL can be regarded as the {\it difference} between the Chern insulator and the strong paired state.    Of course, this scenario occurs for large $U$, far away from any solvable limits.   If correct, then evidently the attractive interaction that binds the electrons and holes into charge $\pm 2e$ pairs must emerge for large (repulsive) $U$.

Our analysis suggests another route to the CSL phase, starting from the trivial insulator with $|E_s|>6t$. Instead of electrons and holes forming charge $\pm 2e$ bosonic pairs via attractive interactions, repulsive interactions may drive binding of $S_z = \pm 1$ excitons $\Phi_{+1}^\dagger = \Psi_{e\uparrow}^\dagger \Psi_{h\uparrow}^\dagger$ and $\Phi_{-1}^\dagger = \Psi_{e\downarrow}^\dagger \Psi_{h\downarrow}^\dagger$. In this case, the analog of (\ref{bosonic nu=2}) using $p=2$ describes a transition from the trivial insulator to the CSL. An important difference between this scenario and the model considered by \cite{kuhlenkamp2024, divic2024a, divic2024b} is that $SU(2)$ spin symmetry
is reduced to $U(1)$ symmetry generated by $S_z$, so it would apply to a model that includes a crystal field or a spin-orbit interaction.   Alternatively, it could be interesting to consider a $SU(2)$ invariant version of the excitonic band inversion problem.

\noindent {\color{blue}\emph{Discussion.}}
In this paper we have studied a model of interacting fermions that undergo a band inversion transition with angular momentum $m$ coupling between the bands.   A composite boson analysis based on a flux attachment scheme dictated by $m$ predicts in the inverted state a strongly correlated fluid of electrons and holes that resembles a Laughlin state with $\sigma_{xy}=(1/m)e^2/h$.   A composite fermion analysis predicts a larger set of possible states that resemble the Jain states.    It remains to be determined whether for $m>1$  (\ref{bhz}) with interactions (or (\ref{bosonic nu=2}) for bosons) exhibits the FEI phase, and if it does whether there is a direct transition to the trivial insulator.   It will be interesting to numerically study this problem, as well as the closely related problem of whether there is a direct transition between the trivial insulator and the chiral spin liquid. This work establishes the paradigm of band inversion combined with flux attachment as a promising organizing principle for these problems.

We thank Stefan Divic and Clemens Kuhlenkamp for helpful discussions.  SG acknowledges funding from the NSF GRFP under Grant No. DGE-1845298.  AS was supported by the CRC TRR-183 on ``Entangled States of Matter", by the Israel Science Foundation, and by the Quantum Program of the Israel Science Foundation.

\onecolumngrid
\begin{center}
  \section{End Matter}
\end{center}
\twocolumngrid

\noindent{{\color{blue}\emph{Appendix A: Composite Boson Action for $p\neq m$.}}}
For completeness, we comment on the general case $p\neq m$ and the gaplessness of the neutral mode. Take $p>m$ without loss of generality. The Lagrangian for the composite boson theory takes the form of (\ref{L}) with the pairing term modified according to (\ref{H pair p><1}). After substituting $\tilde\Phi_{e/h} = \sqrt{n_{e/h}}e^{i\varphi_{e/h}}$, we have
\begin{equation}
    \mathcal{L} = \mathcal{L}_\text{kin.} + \mathcal{L}_\text{pair} + \mathcal{L}_\text{int.} + \frac{1}{m}\mathcal{L}_{CS}[a_\mu]
\end{equation}
\begin{equation}
\begin{split}
    \mathcal{L}_\text{kin.} &= n_e\left(i\partial_\tau\varphi_e-ia_0-iA_0+\frac{1}{2m^*}|\nabla\varphi_e-{\bf a} - {\bf A} |^2\right) \\
    &+ n_h\left(i\partial_\tau\varphi_h+ia_0+iA_0+\frac{1}{2m^*}|\nabla\varphi_h+{\bf a} + {\bf A} |^2\right)
\end{split}
\end{equation}
\begin{equation}
    \mathcal{L}_\text{pair} = \tilde v \sqrt{n_e}e^{-i\varphi_e}(\partial_z - i\bar a)^{p-m}\sqrt{n_h}e^{-i\varphi_h} + \text{h.c.}
\end{equation}
\begin{equation}
    \mathcal{L}_\text{int.} = -E_0(n_e+n_h)+\frac{u}{2}(n_e^2+n_h^2)-wn_en_h
\end{equation}
For $E_0 >0$ and $u>w$, $\mathcal{L}_\text{int}$ favors a ground state with $n_e = n_h = \frac{E_0}{u-w} \equiv \bar{n}$. Changing to neutral/charged variables $n_{e/h} = n_\sigma\pm n_\rho$, $\varphi_{e/h} = \varphi_\sigma\pm\varphi_\rho$, we note that $\varphi_\rho$ appears in $\mathcal{L}_\text{kin.}$ only in the combination $\partial_\mu\varphi_\rho-a_\mu$, and therefore can be eliminated by a gauge transformation on $a_\mu$. Expanding the resulting Lagrangian up to second order in fluctuations around the ground state, we find,
\begin{equation}
\begin{split}
    \mathcal{L}_\text{kin.} &= 2i(\bar n + \delta n_\sigma)\partial_\tau\varphi_\sigma - 2i\delta n_\rho(a_0+A_0) \\
    &+ \frac{\bar n}{m^*}(\nabla\varphi_\sigma)^2+\frac{\bar n}{m^*}({\bf a}+{\bf A})^2
\end{split}
\end{equation}
\begin{equation}
    \mathcal{L}_\text{pair} = \bar n \tilde v e^{-i(\varphi_\sigma+\varphi_\rho)}(\partial_z-i\bar a)^{p-m}e^{-i(\varphi_\sigma-\varphi_\rho)} + \text{h.c.}
\end{equation}
\begin{equation}
    \mathcal{L}_\text{int.} = (\delta n_\sigma)^2(u-w) + (\delta n_\rho)^2(u+w) + \text{const.}
\end{equation}
The term $2i\bar n \partial_\tau\varphi_\sigma$ is a purely topological term with no effect on the local dynamics, so we ignore it for our purposes. Finally, integrating out the fluctuations $\delta n_\sigma$ and $\delta n_\rho$, we arrive at the effective Lagrangian
\begin{equation}
\begin{split}
    \mathcal{L}_\text{eff.} &= \frac{(\partial_\tau\varphi_\sigma)^2}{u-w} + \frac{(a_0+A_0)^2}{u+w} + \frac{\bar n}{m^*}\left((\nabla\varphi_\sigma)^2 + ({\bf a}+{\bf A})^2\right) \\
    &+ \mathcal{L}_\text{pair} + \frac{1}{m}\mathcal{L}_{CS}[a_\mu]
\end{split}
\end{equation}
For $p=m$, $\mathcal{L}_\text{pair}$ reduces to $2 \bar n \tilde v \cos{2\varphi_\sigma}$, recovering (\ref{L sigma rho}). For $p\neq m$, $\mathcal{L}_\text{pair}$ inevitably involves gradient terms. For instance, setting $p=m+2$ and integrating by parts once,
\begin{equation}
\begin{split}
    \mathcal{L}_\text{pair} &= -\bar n \tilde v e^{-i2\varphi_\sigma}\partial_z(\varphi_\sigma+\varphi_\rho)\partial_z(\varphi_\sigma-\varphi_\rho) + \text{h.c.} \\
    &= \bar n \tilde v e^{-i2\varphi_\sigma}[(\partial_z\varphi_\rho)^2-(\partial_z\varphi_\sigma)^2] + \text{h.c.}
\end{split}
\end{equation}

\noindent{{\color{blue}\emph{Appendix B: Comparison of Energy Scales between Different Ground States.}}}
We are interested in identifying a parameter regime in the single-particle Hamiltonian (\ref{bhz}) that is likely to favor the ground state of the composite boson theory (\ref{L sigma rho}) in the presence of realistic interactions. This is analogous to efforts in the FCI community to identify features of flat Chern bands (e.g. ``ideal quantum geometry") that favor FCIs as interacting ground states. Of course, the answer to this question depends on the precise form of the interactions, which we do not specify. But we can still make a rough comparison between the $\sigma_{xy}h/e^2= m$ and $\sigma_{xy}h/e^2 = 1/m$ states as follows.

In the absence of interactions, the ground state of $\mathcal{H}_0$ in Eq. (\ref{bhz}) is a semimetal with Fermi wavevector $k_F \sim E_0^{-1/2}$. Turning on $\mathcal{H}_\text{pair}$ with no additional interactions stabilizes a Chern insulating state with $\mathcal{C} = m$, opening a gap $\Delta \sim v k_F^m$ at the Fermi surface. We can estimate the energy gain (difference in ground state energy between $\mathcal{H}_0$ and $\mathcal{H}_0 + \mathcal{H}_\text{pair}$) as
\begin{equation}
\begin{split}
    \delta E_\text{CI} &= N_0\int_0^\Lambda d\epsilon\left(\epsilon-\sqrt{\epsilon^2+\Delta^2}\right)
    \overset{\Lambda \gg \Delta}{\sim} -\frac{N_0\Delta^2}{2}\ln{\frac{\Lambda}{\Delta}}
\end{split} \label{delta E CI}
\end{equation}
where $N_0$ is the density of states at the Fermi surface and $\Lambda$ is an energy cutoff (in the last expression, we absorb unimportant constants into a redefinition of $\Lambda$).

Consider now the interacting theory, which we describe in terms of the $p=m$ composite boson theory in (\ref{L sigma rho}). Starting from the gapless $v = 0$ state, turning on $v$ opens a many-body gap with an energy gain $\delta E_\text{FEI} \sim -\bar n v/d^m$, where $\bar n$ denoting the average density of electrons and holes. Comparing this with (\ref{delta E CI}) with the same density of electrons and holes ($N_0 = \bar n / E_0$), we have
\begin{equation}
    \frac{\delta E_\text{FEI}}{\delta E_\text{CI}} \sim \frac{E_0}{\Delta}\frac{1}{(k_F d)^m} 
    .\label{ratio}
\end{equation}
Note that this is \textit{not} an absolute energetic comparison between the Chern insulating state and the FEI for some interacting Hamiltonian. Rather, this compares how each state takes advantage of the pairing term to lower its energy relative to the $v=0$ ground state. Nonetheless, the parameter regime that makes the ratio (\ref{ratio}) large is a helpful starting point for further investigation of interacting Hamiltonians. In particular, when $\Delta$ is small compared to $E_0$ and the Berry curvature is spread over a range $1/d$ that is large compared to $k_F$, in the Chern insulating phase only a small number of electrons/holes near the Fermi surface lower their energy via the pairing term. In contrast, the FEI allows all electrons/holes with momentum $k\lesssim 1/d$ to lower their energy via the pairing term by locking the neutral phase mode of the composite bosons.

It is also worth noting that like Laughlin's wavefunction, the wavefunction in Eq. (\ref{wavefunction}) keeps like charges apart, and will therefore be favored by short-ranged repulsive interactions.

\noindent{{\color{blue}\emph{Appendix C: $K$-Matrix Description.}}} The two-component nature of the wavefunction (\ref{wavefunction}) suggests a description of the $\sigma_{xy} = (1/m)e^2/h$ FEI states in terms of a multi-component Chern-Simons action $\exp(iS) = \exp(i\int dt\,d^2x\, \mathcal{L}$) with
\begin{equation}
    \mathcal{L} = \frac{1}{4\pi}\sum_{IJ} K_{IJ} \epsilon_{\mu\nu\lambda}a_{I,\mu}\partial_\nu a_{J,\lambda} + \frac{1}{2\pi}\sum_I t_I\epsilon_{\mu\nu\lambda}A_\mu\partial_\nu a_{I,\lambda}
\end{equation}
and with $K$-matrix and charge vector given by,
\begin{equation}
    K =
    \begin{pmatrix}
        m & -m \\
        -m & m
    \end{pmatrix}
    , ~~~
    t = \begin{pmatrix}
        1 \\
        -1
    \end{pmatrix}
    .
\end{equation}
The two components correspond to electrons and holes. As noted in \cite{Hu2018}, this is related to the Halperin $(m,m,-m)$ bilayer state, or the $(m,m,m)$ state after a particle-hole transformation in one layer. The $K$-matrix is degenerate, and can be transformed via a $SL(2,\mathbb{Z})$ transformation to
\begin{equation}
    K' =  \Lambda K \Lambda^T =   \begin{pmatrix}
        m & 0 \\
        0 & 0
    \end{pmatrix}
    , ~~~
    t' = \Lambda t =  \begin{pmatrix}
        1 \\
       0
    \end{pmatrix}
    ,
\
\end{equation}
with
\begin{equation}
    \Lambda = \begin{pmatrix} 1 & 0 \\ 1 & 1 \end{pmatrix} . \ 
\end{equation}
The zero eigenvalue corresponds to the neutral sector that is gapped by the pairing term in the $p=m$ composite boson theory (\ref{L sigma rho}), and we are left with an effective single component theory described by $K=m$. In analogy to the quantum Hall ferromagnet, where spontaneous breaking of a spin-rotation symmetry leads to a spin-polarized quantum Hall state, here it is the breaking of the independent number conservation of electrons and holes (enabled by $v\neq 0$) that leads to an effective single-component FEI state.

\end{document}